# An Evaluation of Level of Detail Degradation in Head-Mounted Display Peripheries

**Benjamin Watson, Neff Walker & Larry F. Hodges**

Graphics, Visualization & Usability Center

Georgia Institute of Technology

801 Atlantic Drive, Atlanta, GA 30332-0280, USA

**Martin Reddy**

Department of Computer Science

University of Edinburgh

Edinburgh, United Kingdom

**Abstract.** A paradigm for the design of systems that manage level of detail in virtual environments is proposed. As an example of the prototyping step in this paradigm, a user study was performed to evaluate the effectiveness of high detail insets used with head-mounted displays. Ten subjects were given a simple search task that required the location and identification of a single target object. All subjects used seven different displays (the independent variable), varying in inset size and peripheral detail, to perform this task. Frame rate, target location, subject input method, and order of display use were all controlled. Primary dependent measures were search time on trials with correct identification, and the percentage of all trials correctly identified. ANOVAs of the results showed that insetless, high detail displays did not lead to significantly different search times or accuracies than displays with insets. In fact, only the insetless, low detail display returned significantly different results. Further research is being performed to examine the effect of varying task complexity, inset size, and level of detail.

## 1. Introduction

As researchers attempt to broaden the range of applications for virtual environments (VE) technology, the complexity of the models they are using is increasing. Unfortunately, as the complexity of a model increases, so does the time it takes to display a view of it. VE researchers are having difficulty displaying their models at interactive rates.

Many have identified this "frame rate" problem as one of the most pressing facing the VE community (Bryson, 1993; Furness, 1991; NSF, 1992; Van Dam, 1993). Foremost among the proposed solutions to this problem is the idea of varying "level of detail" (LOD). This phrase refers to model and rendering complexity, which can be managed on the fly to ensure that VEs are rendered at some minimal frame rate (Heckbert & Garland, 1994; Helman, 1993).

In the design of systems that manage LOD in this manner, careful consideration should be given not only to the computational costs of graphical rendering techniques, but also to their perceptual costs. If two rendering techniques make similar demands on the graphics engine, but the use of one of the techniques makes only a minimal contribution to perceptual fidelity or presence (Heeter, 1992; Sheridan, 1992; Slater & Usoh, 1993a; Zeltzer, 1992), then that technique should be the first to go in the effort to maintain frame rate. Any other decision would clearly be wasteful. This leads us to propose the following paradigm for the design of LOD management systems:

1) The human perceptual system is examined, and those characteristics that might be exploited by an LOD management system are identified.

2) A prototype of the management system is then developed, and the effectiveness of the techniques it uses verified and refined.

3) Assuming the prototype works, the actual system is implemented.

In this paper, we identify one characteristic of human vision that might be exploited in an LOD management system. We then prototype and test such a system. In a companion paper (Reddy et al., 1996), we perform a more general evaluation of the human visual system, and propose an LOD management system that leverages several of its characteristics.

**2. Peripheral LOD Degradation**

Perception is not uniform across the visual field. Many measures of ability to perceive detail, including visual acuity, contrast sensitivity, stereo acuity, and temporal sensitivity vary with retinal eccentricity (Bishop, 1986; Graham, 1989; Spillman & Werner, 1990). In contrast, VEs often spread complexity and computation evenly across the raster display. This suggests the possibility of a computationally and perceptually efficient peripherally degraded display containing a central, high detail inset, corresponding to the perceptual characteristics of the foveal area of the retina; as well as a surrounding, simpler periphery, corresponding to the perceptual characteristics of the peripheral area of the retina. We call this technique peripheral LOD degradation.

Howlett (1992), Slater & Usoh (1993b) and Yoshida, Rolland & Reif (1995) have all proposed peripherally degraded display systems, but the focus in their work was not improving frame rate. Both Funkhouser & Séquin (1993) and Maciel & Shirley (1995) have implemented systems that use peripheral degradation. However, neither of these systems has undergone rigorous usability testing.

There are many rendering techniques that might be used to vary image complexity, including using geometric models of varying degrees of accuracy (DeRose & Lounsberry, 1993; Rossignac & Borrel, 1992; Turk, 1992; Varshney et al., 1995), lighting models of differing levels of realism, and textures and graphics windows of differing resolution (Maciel & Shirley, 1995). Many researchers have compared the relative importance of these and other graphical display techniques (Atherton & Caporeal, 1985; Barfield, Sandford & Foley, 1988; Booth et al., 1987; Todd & Mingola, 1983). In general, these studies showed significant effects on performance when image complexity is varied. However, in most cases a point of diminishing returns was reached, beyond which additional image complexity and computation produced insignificant performance improvement.

We chose to prototype and evaluate peripheral degradation with the use of the computationally simplest of these techniques, varying window resolution. Moreover, because currently available eye tracking technology is unwieldy and expensive, we worked under the assumption that head-tracking alone would allow effective peripheral degradation. In studies that examined varying resolution without peripheral degradation, Booth et al. (1987) found decreasing subjective preferences and increasing task performance times as window resolution was decreased. Smets & Overbeeke (1995) showed that frame rate is more important than resolution for many tasks.

We believe that peripheral LOD degradation, when implemented in VEs, will result in minimal perceptual loss and significant computational gain. The computational portion of this assertion has already been examined in (Funkhouser & Séquin, 1993; Maciel & Shirley, 1995). In our study, we attempted to verify the perceptual portion of this assertion by measuring subject performance time and accuracy while peripheral degradation was used with varying inset sizes and levels of peripheral detail.

We chose search as the benchmark task because we felt that a search task formed a worst case for a peripherally degraded display. Treisman (Treisman & Gelade, 1980; Treisman, 1986) has studied the factors of human performance in two dimensional search tasks, with the display lying entirely within the subject's field of view and multiple distracting objects present. The resulting model identified several types of visual features that are processed at an early stage of the perceptual process. If the sought after object is the only that contains a given feature, that portion "pops out", and the search concludes quickly. If the sought after object is uniquely defined only by a certain combination of features, no "pop out" occurs. Our study differs from Treisman's in that the display was three dimensional, and extended beyond the edges of the subject's visual field. Furthermore, no distractors were used. Nevertheless, Treisman's work suggests that when peripheral fidelity is being reduced, care should be taken to preserve the perceptual features important for the task being performed. In general, these features should become apparent in step 2 of our LOD management paradigm above.

Because visual acuity and sensitivity decrease with eccentricity, we expected that loss of detail would have less impact on subject performance time and accuracy when peripheral degradation was used than when an evenly degraded, low detail display was used. We anticipated that use of the evenly degraded, high detail display, which makes full use of display resolution, would result in the lowest subject performance times and highest accuracies.

We had no way of predicting either the optimal visual extent of the high detail inset in the peripherally degraded display, or the ideal LODs in the high detail inset and the periphery of the display. We investigated the effect of changing these variables by using high detail insets of two different sizes, and three different resolution levels.

### 3. Experimental Method

Ten college students, including both graduates and undergraduates, participated in the study. All subjects were experienced with virtual reality and head-mounted displays, and exhibited at least average corrected vision in testing.

Subjects wore a Virtual Research Flight Helmet (Robinett & Rolland, 1992; Rolland & Hopkins, 1993). The Virtual Research Flight Helmet mounts two color LCD displays on the user's head, each with vertical field of view of 58.4 degrees, and a horizontal FOV of 75.3 degrees. Each LCD contains an array of 208 x 139 color triads, with an average horizontal resolution of 21.74 arcmins. Although this maximal resolution is somewhat coarse in comparison to some newer displays, it represents a level of detail that will often be reached as detail is globally degraded in more complex environments. The Flight Helmet weighs 3.7 pounds, and takes two NTSC signals as input. We used the Flight Helmet in a monoscopic mode by sending the same image to each of the video inputs, and mounting plastic fresnel lenses on the HMD optics to remove interocular disparity.

The motion of a subject's head in the Flight Helmet was tracked with the Polhemus Isotrak II 3D tracking hardware. The monoscopic images sent to the Flight Helmet were generated by a Silicon Graphics Onyx Reality

Engine II, or on occasion a Silicon Graphics Reality Engine, using the gl graphics library and the SVE virtual environments library (Kessler, 1993). Silicon Graphics own scan converting hardware and software was used to convert these images into an NTSC signal. Subjects used a plastic mouse shaped like a pistol grip to respond to the experimental environment. The mouse had two buttons for the thumb mounted on top, and one button for the index finger mounted on the front. The mouse was not tracked. When using the experimental environment, subjects stood inside a 4x4 platform raised six inches and surrounded by a 3 foot railing. This kept subjects within four feet of the Isotrak transmitter.

The virtual experimental environment consisted of a floor, indicated by a grid of white lines on black (see figure 1). The background above the floor was also black. Users could see no analogue of themselves in the virtual experimental environment. Between trials, subjects would focus on a home object. The home object was a flat, white panel, labeled with a red bullseye design (see figure 2).

*figures 1 and 2*

During each trial, subjects would search for a white target object, onto which a red uppercase letter (from A to J) or number (from 0 to 9) had been textured. The number 0 was distinguished from the letter O with a slash, the number 1 from the letter I with a downward sloping serif on the top. The target always appeared at the same virtual distance, and subtended a horizontal visual angle of approximately 13 degrees.

Each experimental trial consisted of a single search task. After focusing on the home object, subjects pressed a button to begin the task. After a random (between .1 and .8 seconds) delay, the home object disappeared, and a single target object appeared outside the subject's initial view. Subjects located the target object and pressed one of two buttons to indicate if the object was labeled with a number or a letter. The target object then disappeared, and the home object reappeared. At the same time, onscreen feedback was provided indicating if the correct button had been pressed, and the number of seconds their search required. When the subjects had again focused on the home object and pressed the appropriate button, a new search trial began. The onscreen feedback was present for at most five seconds, and disappeared in any case as soon as the next trial began.

Subjects performed the search task with seven different display types. Each of these display types was a combination of the two main independent variables: peripheral resolution and high LOD inset size. Each of these variables was varied within subjects at three possible levels. The fine level of peripheral resolution of the image scanned into the HMD was 25% of NTSC: 320 x 240 pixels. Medium resolution was 9% of NTSC: 192 x 144.

Coarse resolution was only 1% of NTSC: 64 x 48. Note that at all of these resolution levels, the image remained at a constant size.

The high detail inset was rectangular, and was always presented at the fine level of resolution. The largest inset occupied one quarter of the display space (and had half the complete display's height and width). The smallest inset occupied only 9% of the available display space (with only 30% of the complete display's height and width). At the third inset size level, no inset was present. The size of the image generated for texturing into the inset was adjusted to ensure constant pixel size corresponding to the fine level of resolution. The seven different combinations of these variable levels used to create the seven different display types are listed in Table 1.

Several other variables were controlled. Since the maximum frame rates possible in the different display types varied, the minimum of these possible frame rate maximums was selected and used as an upper bound for the frame rates in all display types. This frame rate was 12 Hz, and the average resulting frame rate was 11.96 Hz, with an average standard deviation of 0.12 Hz. Target object location was controlled through the use of nine regions located around the subject, as illustrated in figures 3 and 4. These regions were of equal area, and no regions were located above, below, or in front of the subject when in home position. Care was taken so that targets would not overlap into neighboring regions. The letters or numbers on the target objects were randomly chosen. The button subjects used to indicated the presence of a letter (or number) was randomly varied and counterbalanced between subjects. Subjects worked with one display type until all trials with that display type were complete; however, the order in which different display types were presented was randomly varied between subjects, and counterbalanced so that no display was presented as the $n$th display three times, and so that no two display sequence was presented four times. Subjects were not permitted to end a trial unless the target object had actually been displayed on the HMD.

*figures 3 and 4*

Display type was varied within subjects, with each subject using all seven displays. Each subject continued working with a display until 90 target objects were correctly identified (the letter or number was correctly identified), for a total of 630 correct trials over all display types. For the trials with a given type, the target object was located in each of the regions for 10 correct trials. The order of these region locations was random. As noted above, the main dependent variables were search time on correct trials, and target identification accuracy. Search time of incorrect trials was not included because we felt that a significant portion of these trials would result from accidental button presses.

Before beginning the experiment, subjects read a two page introduction to the purpose of the experiment and its procedure. This explained, among other things, that subjects were permitted to pause between any two trials if they required a rest. It also made subjects aware that they would be ranked by search time and accuracy, with the subject with the best cumulative ranking receiving $50 after the completion of the experiment.

Most subjects participated in three experimental sessions. Subjects were required to complete all trials with a given display before ending a session. Before beginning the first session, subjects were allowed 20 search trials as practice. At the beginning of each session, subjects were presented in sequence with five target objects located directly in the center of their view. This allowed subjects to reacquaint themselves with their button configuration. Before searching with a new display, subjects were allowed five practice search trials, so that they might familiarize themselves with the new display.

### 4. Results

The two primary dependent measures we used in our analysis of the results were accuracy and search time. Accuracy was defined as the percentage of correct search trials out of the total number of searches in a condition. Search time was the average time to find a target and make a correct identification.

Initially we performed a display type (7) by location (9) by button configuration (2) analysis of variance (ANOVA) on both accuracy and search time to discover if subject button configuration interacted with the other two independent variables. This analysis revealed no main effect or signification interaction. For all further analyses, this factor was collapsed over.

We made two primary analyses. First we ran a display type by location ANOVA on accuracy. This analysis revealed only a significant effect of display type ($F(6,64) = 38.60$, $p < .001$). We used bonferroni pair-wise comparisons to determine which display types lead to different accuracies, using an adjusted probability level of 0.05. These comparisons revealed that all display types lead to significantly higher accuracies than the insetless, coarse resolution display condition (see table 1 and figures 5 and 6).

*table 1*

We next ran a display type by location ANOVA on search time. This analysis revealed a significant main effect of display type ($F(6,54) = 5.39$, $p < .001$) and a main effect of location ($F(8,72) = 13.80$, $p < .001$). The interaction between display type and location was not significant. Bonferroni pair-wise comparisons for display type revealed that the insetless, fine resolution display had significantly shorter search time than the insetless, low resolution display. Furthermore, the insetted displays were not significantly different from the insetless, fine resolution display (see figure 6). The pair-wise comparisons for the main effect of location revealed that searches for targets in the three upper regions (7, 8 and 9 in figure 4) took significantly longer than searches for targets in other regions.

In an effort to investigate possible differences in accuracy and search time due to inset size and peripheral resolution, we ran a three-way ANOVA on these two variables. The factors in the analysis were peripheral resolution (2 levels) by inset size (2 levels) by region. Only location effects proved to be significant.

*figures 5 and 6*

One might expect to see a bimodal distribution in search time, with search times being short if the target was located in the direction of initial head motion, longer otherwise. There was no such evidence. In fact, since subjects could not see the entire vertical extent of the area to search, most subjects used an elliptical search pattern, and the distribution of search times seemed to indicate that the target was generally found halfway through a single rotation around this ellipse.

## 5. Discussion

Results indicated that peripheral LOD degradation can be a very useful compromise. The display type with the lowest LOD -- an insetless, low resolution display -- was significantly worse than any insetted display, even if the insetted display used low peripheral resolution. At the same time, the display type with the highest LOD -- an insetless, high resolution display -- was not significantly better than insetted displays of any type. The fact that these results were achieved without eye tracking is particularly interesting, and suggests that eye tracking may be of little importance in HMDs when the high LOD inset is not extremely small.

Unfortunately, we were not able to draw any conclusions about optimal inset size or LOD (in resolution) from our results. A trend relating increased LOD (in resolution) to decreasing search times did exist (figure 6), but was not significant. We suspect that this is due to the nature of the experimental task. We hypothesize that search in this experiment involved two phases: motion to the area of interest, followed by closer examination. All peripheral resolutions proved adequate for the first phase. However, since even the smallest inset could contain most of the target object, peripheral resolution proved irrelevant in the second phase, and no significant effects of inset size or resolution resulted.

## 6. Future Work

Our hypothesis suggests that smaller inset sizes and larger areas of interest should increase search times. If an area of interest that fits into a screen but not into an inset is viewed, head motion rather than eye motion will be required during the identification phase, and search times will increase. We plan on testing this hypothesis in a follow-up experiment. The experiment will introduce a new variable, clusteredness, which measures the size of the area of interest. Through the use of this variable and a new, small inset size level, we hope to obtain some sense of the effect of varying inset size and peripheral resolution.

**Captions**

**Figure 1:** View of the target and floor in the experimental environment. Peripheral resolution is coarse, inset size is large.

**Figure 2:** View of the home object in the insetless, coarse resolution display in the experimental virtual environment. Note the feedback for correct identification and search time.

**Figure 3:** The search space around the subject was divided into nine regions. Targets were not located directly above or below the subject, or in view in home position.

**Figure 4:** Top down schematic view of the user surrounded by the search regions. Here the regions are numbered in counterclockwise and low to high elevation order. The home object is at the top.

**Figure 5:** Average accuracy for each display in percentage of correct identifications. Displays are grouped by inset size and peripheral resolution.

**Figure 6:** Average search times for each display type in seconds. Displays are grouped by inset size and peripheral resolution.

**Table 1:** The seven display types. Performance with a display is shown as means and standard deviations for both accuracy (pctg correct) and search time (seconds on correct searches).

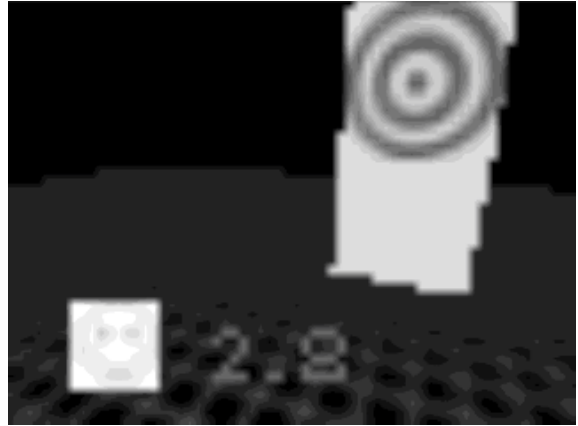

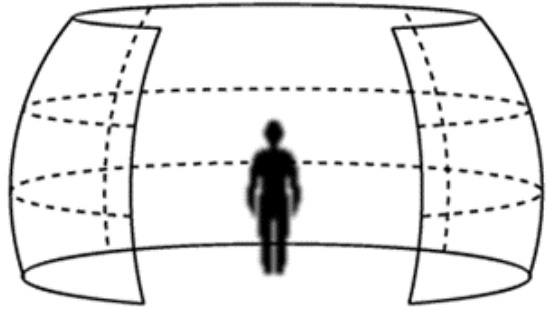

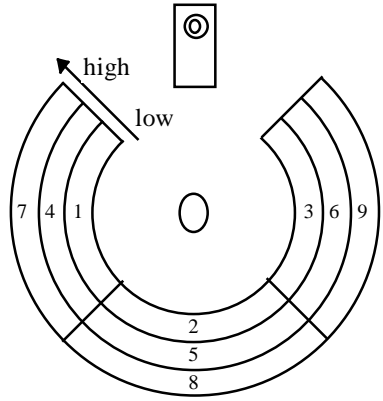

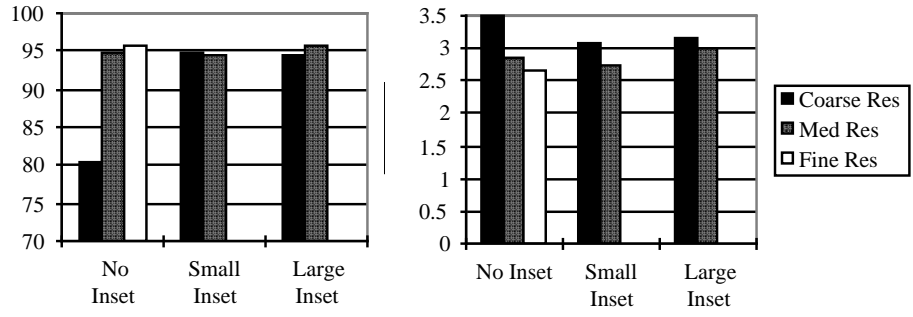

**Table 1:**

| Display Type | Pixels in Display | Accuracy (pctg) | | Srch Time (secs) | |
|---|---|---|---|---|---|
| | | Mean | Std Dv | Mean | Std Dv |
| No inset, fine res | 76800 | 95.7% | 6.74 | 2.652 | 0.793 |
| No inset, med res | 27648 | 94.8% | 6.84 | 2.863 | 0.878 |
| No inset, coarse res | 3072 | 80.3% | 12.85 | 3.490 | 0.993 |
| Lg inset, med res periph | 45216 | 95.9% | 6.11 | 2.986 | 0.825 |
| Lg inset, coarse res periph | 29952 | 94.6% | 6.90 | 3.146 | 0.930 |
| Sm inset, med res periph | 35252 | 94.5% | 6.90 | 2.727 | 0.790 |
| Sm inset, coarse res periph | 15084 | 94.9% | 8.08 | 3.097 | 1.033 |